\documentclass[preprints,article,accept,moreauthors,pdftex]{Definitions/mdpi}

\firstpage{1} 
\pubvolume{1}
\issuenum{1}
\articlenumber{0}
\pubyear{2021}
\copyrightyear{2020}
\datereceived{} 
\dateaccepted{} 
\datepublished{} 
\hreflink{https://doi.org/} 

\Title{Properties of Faint X-ray Activity of XTE J1908+094 in 2019}

\Author{Debjit Chatterjee $^{1,2}$*\orcidA{}, Arghajit Jana $^{3}$\orcidB{}, Kaushik Chatterjee $^{2}$\orcidC{}, Riya Bhowmick $^{2}$\orcidD{}, Sujoy Kumar Nath $^{2}$\orcidE{}, Sandip K. Chakrabarti $^{2}$\orcidF{}, A. Mangalam $^{1}$\orcidG{} and Dipak Debnath $^{2}$\orcidH{}}

\address{%
$^{1}$ \quad Indian Institute of Astrophysics, Koramangala, Bangalore, 560034, India\\
$^{2}$ \quad Indian Centre for Space Physics, 43 Chalantika, Garia St. Rd., Kolkata, 700084, India\\
$^{3}$ \quad Physical Research Laboratory, Navrangpura, Ahmedabad 380009, India}

\corres{debjit.chatterjee@iiap.res.in}

\abstract{We study the properties of the faint X-ray activity of Galactic transient black hole candidate XTE~J1908+094 during its 2019 outburst. 
Here, we report the results of detailed spectral and temporal analysis during this outburst using observations from {\it Nuclear Spectroscopic 
Telescope Array (NuSTAR)}. We have not observed any quasi-periodic-oscillations (QPOs) in the power density spectrum (PDS). 
The spectral study suggests that the source remained in the softer (more precisely in soft-intermediate) spectral state during this 
short period of the X-ray activity. We notice a faint but broad Fe K$\alpha$ emission line at around 6.5 keV. We also estimate 
the probable mass of the black hole to be $6.5^{+0.5}_{-0.7}~M_\odot$ with 90\% confidence.
    }


\begin{document}


\section{Introduction}

Black hole transients (BHTs) are fascinating objects to study. After a long period of quiescence, they show a sudden outburst. 
The sudden enhancement of viscosity at the piling radius could trigger the outburst \citep{SKC96,SKC19,Mondal17}. The spectral and 
temporal properties of the source change during the outburst and evolves through the hard state (HS), hard intermediate state (HIMS), 
soft intermediate state (SIMS), and soft state (SS) \citep{RM06,MR09,DD13}. Evolution of the state  can be seen through 
hardness intensity diagram (HID) or "q" diagram \citep{Belloni05,Belloni10} and accretion rate ratio intensity diagram \citep[ARRID;][]{AJ16,KC20}. 
A "failed" outburst is also a commonly known event where the source does not enter the softer spectral states \citep{Tetarenko16,Garcia19,DC19}. 
In case of "failed" outbursts, sources do not follow the standard state transition or "q" diagram.
The spectral shape
varies in different spectral states mainly due to the relative contribution of thermal component \citep{SS73,NT73} and non-thermal component \citep{ST80,ST85}. 

In Two-Component Advective Flow (TCAF) solution \citep{CT95,SKC97}, the accretion flow consists of two components: high viscous 
Keplerian flow with high angular momentum and low viscous sub-Keplerian flow with low angular momentum. The sub-Keplerian flow 
moves towards the black hole almost radially, and it almost stops at the centrifugal barrier and forms an axisymmetric shock \citep{SKC90}. 
The matter gets puffed up beyond the shock and creates a hot electron cloud or Compton corona. This corona is called the CENtrifugal 
pressure supported BOundary Layer or CENBOL. This region intercepts the soft photons coming from the Keplerian disk and emits 
high-energy photons through inverse-Comptonization. In this way, TCAF can self-consistently explain the accretion dynamics around 
an accreting black hole. The oscillation of the same shock can explain the observed low-frequency quasi-periodic-oscillations (LFQPOs) \citep{MSC96,Nandi12}. The CENBOL is also considered to be the base of jets and outflows \citep{SKC99}. To get an estimation of the 
physical flow parameters directly from the spectral fit, TCAF model was implemented as an additive table model in {\tt XSPEC} 
\citep{DD14,DD15}. From spectral fit with the model, we obtain two accretion rate parameters, namely, the Keplerian disk rate 
($\dot{m}_d$) in Eddington rate ($\dot{M}_{Edd}$), the sub-Keplerian halo rate ($\dot{m}_h$) in Eddington rate ($\dot{M}_{Edd}$); 
two shock or Compton cloud parameters- shock location ($X_s$) in Schwarzschild radius ($r_s$), compression ratio ($R$), 
which is the ratio between post-shock and pre-shock matter densities ($R=\rho_+/\rho_-$). One also gets the best-fitted value of the mass 
of the black hole ($M_{BH}$ in $M_\odot$) and a normalization parameter from each spectral fit. If the mass of the black hole is well known, 
it could be kept constant during the spectral fitting.

The Galactic black hole (GBH) candidate XTE~J1908+094 was discovered on 2002 February 21 by the Proportional Counter Array (PCA) on-board 
Rossi X-ray Timing Explorer ({\it RXTE}) \citep{Woods02}. The source spectrum was fitted with an absorbed power-law with photon 
index ($\Gamma$) of $1.55$. The power density spectrum did not show any pulsations and exhibited a flat spectrum between 1 mHz and 0.1 Hz. 
A broad quasi-periodic-oscillation peak at 1 Hz was observed with a power-law break which continued to 4 Hz. A high energy cutoff at 
$\sim$100 keV was also observed using {\it BeppoSAX} \citep{Feroci02}. The source was suggested to be a black hole (BH) candidate from its 
spectral and timing properties \citep{Woods02,Feroci02,int02}. High interstellar absorption (column density, $N_H\sim2.5\times10^{22}~cm^{-2}$) 
was reported while fitting the spectrum with a multi-color blackbody, a Comptonization and a broad emission line \citep{int02}. 
The dimensionless spin parameter was also measured to be 0.75 from the broadening of Fe K$\alpha$ line \citep{Miller09}. 

A radio counterpart was discovered at R.A.=$19^h08^m53.^s07$, DEC.=+$09^\circ 23'05''.0$ by {\it Very Large Array (VLA)} \citep{Rupen02} 
which was also consistent with the Chandra observation \citep{Jonker04}. Two possible near-infrared (NIR) counterparts were detected 
\citep{Chaty02,Chaty06} - one of them is indicated as an intermediate/late-type (A-K) main-sequence companion, while the other is suggested 
to be a late-type (later than K) main-sequence secondary star \citep{Chaty06}. After two similar outbursts in 2002 and early 2003, 
XTE~J1908+094 went through another outburst on 2013 October 26 \citep{Krimm13, Miller-Jones13, Rushton13, Negoro13, Coriat13}. 
The 2013 outburst was well studied in X-rays by {\it Swift}, {\it NuSTAR} \citep{Tao15,Zhang15}. Although a relativistic broadening of Fe-K$\alpha$ line was observed, the spin of the BH could not be constrained due to data quality. A disk reflection contribution was also observed. The source was in the high/soft state during the {\it NuSTAR} observation. A flare was observed during the studied period. The flare was suggested to be related to the relativistic jet activity. Changes in the corona could be the reason for the flare. Multi-frequency radio and X-ray observation, and radio polarimetry with VLA and AMI-LA during the entire 2013 outburst was done \citep{Curran15}. The source followed the standard hardness-intensity diagram during the outburst. The common behavior of radio jets was also observed that changes from compact to discrete as the state transits from hard to soft. From the VLBI monitoring of XTE J1908+094, a lateral expansion of resolved, asymmetric jet knots was noticed, which was ejected following the hard to soft state transition \citep{Rushton17}. The knots are suggested to be the working surface where the ejected materials interacted with the surrounding dense interstellar medium. An external shock formed in this region causes the acceleration of particles which subsequently diffused outwards over time.

XTE~J1908+094 recently showed a "faint" X-ray activity on 2019 April 1 \citep{Rodriguez19}. The source spectrum was fitted with a power-law model 
with photon index, $\Gamma$ = 2.3. AMI-LA 15.5 GHz observation on 2019 April 5, detected the radio counterpart of XTE~J1908+094 \citep{Williams19}. 
The obtained spectra on 2019 April 4 from the photon counting mode of Swift also yield a soft spectrum with high absorption \citep{Miller19}. 
The source was suggested to be in the soft spectral state while using NICER data on 2019 April 6 and 9 \citep{Ludlam19}.

In this {\it paper}. we have studied the timing and spectral properties of XTE~J1908+094 during its X-ray activity on 2019 April 10 using 
{\it NuSTAR} observation of $\sim$40~ks. The {\it paper} is organized in the following way. In \S 2, we discuss the observations and the 
data analysis procedure. In \S 3, we present the temporal and spectral results of our analysis. In \S 4, we carry out the discussion based 
on our results.

\section{Observation and data analysis}

We processed the {\it NuSTAR} observation Id 90501317002\footnote{\url{https://heasarc.gsfc.nasa.gov/cgi-bin/W3Browse/w3browse.pl}} 
(Date: 2019 April 10) using the {\tt nupipeline} command of NuSTAR Data Analysis Software (NuSTARDAS) version 1.8.0 
\footnote{\url{https://heasarc.gsfc.nasa.gov/docs/nustar/analysis/}} and with the Calibration database (CALDB) version 
1.0.2\footnote{\url{http://heasarc.gsfc.nasa.gov/FTP/caldb/data/nustar/fpm/}}. The spectra and light curves of the source were extracted 
from the FPMA detectors using a 60$''$ circle centered at the position of XTE~J1908+094. The background region was selected carefully 
since the source could have contamination from the nearby bright source GRS~1915+105. The background was chosen as a 60$''$ radius circle 
as far away from the XTE~J1908+094. The background count rate is less than 2\% of the source count rate, indicating that the source is 
still dominant. We also divided the total $\sim$40~ks data in three segments of $\sim$16~ks, $\sim$10~ks, and $\sim$12~ks for detailed 
study. We used {\tt XSELECT} command {\tt filter time} for this. The spectra and light curves were then generated using {\tt nuproducts}. 
The spectra were re-binned to have 20 counts/sec using {\tt grppha}. The light curves were binned with 100 sec time resolution.

For spectral analysis, we used both phenomenological (combined {\tt diskbb}, {\tt powerlaw} models) and physical (TCAF based {\tt fits} file 
as an additive table model) model in {\tt XSPEC} version 12.10.1. The hydrogen column density ($N_H$) was fixed at 2.5$\times$10$^{22}$~cm$^{-2}$ 
\citep{int02}. The multiplicative model, {\tt Tbabs} was used as an absorption table model considering the {\tt vern} scattering cross-sections 
\citep{Verner96} and {\tt wilm} abundances \citep{Wilms00}.  

\section{Results}
\begin{figure}[ht]
    \centering
    \includegraphics[width=10.0cm,keepaspectratio=true]{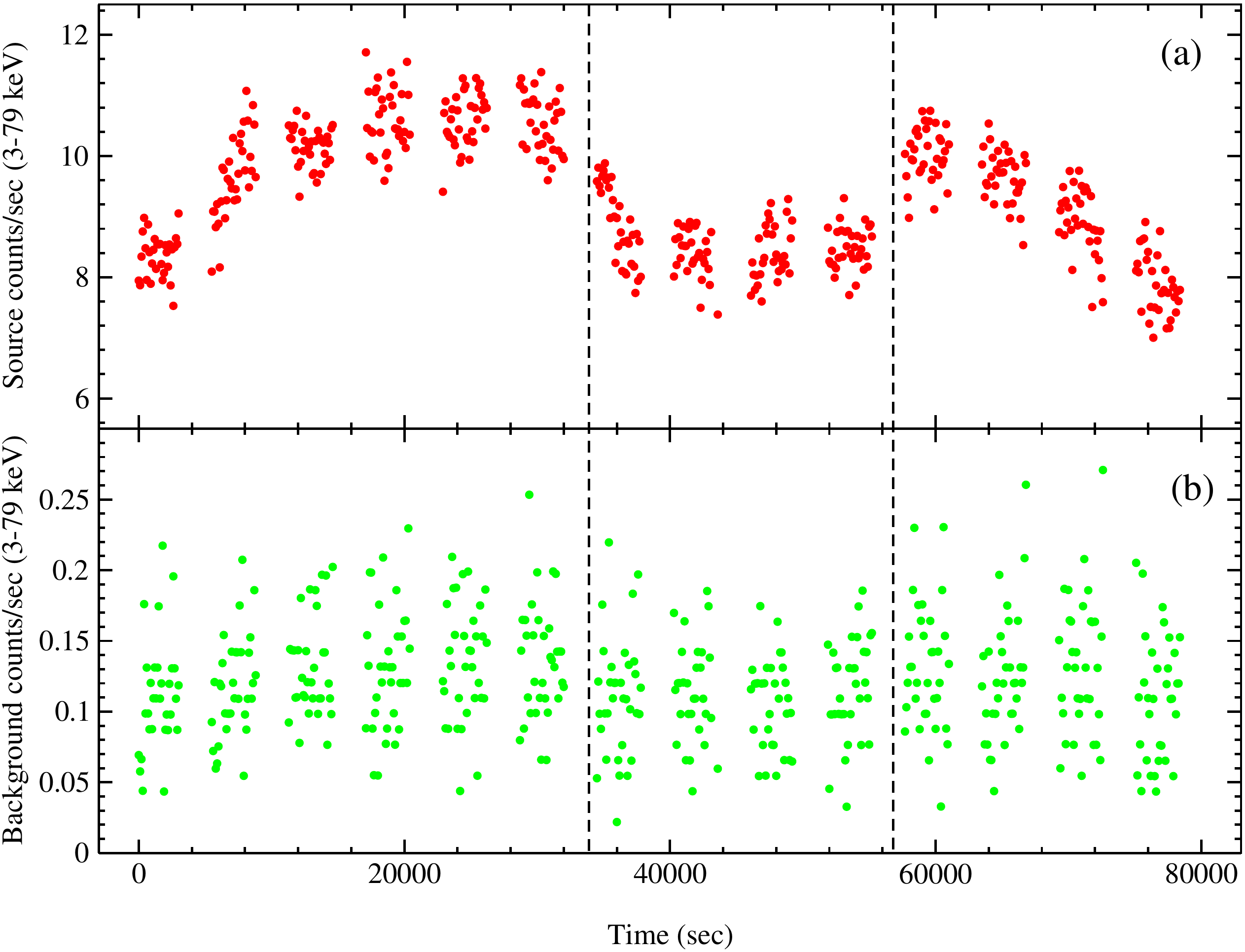}
    \caption{Variation of count rates of XTE~J1908+094 from {\it NuSTAR} observation on 2019 April 10. The upper panel (a) shows the count rate 
    for the source and the lower panel (b) shows the count rate of the background. The background counts is less than 2\% of the source counts.}
    \label{src-bkg}
\end{figure}

We studied the 2019 X-ray activity of BHC XTE~J1908+094 using {\it NuSTAR} observation. The source is close to a bright BHC GRS~1915+105. 
To verify the presence of any contamination due to GRS~1915+105, we studied the source and background count rate variation of XTE~J1908+094 
(Fig.~\ref{src-bkg}). The background count rate was found to be less than 2\% of the source count rate. So, we conclude that the X-ray 
activity of XTE~J1908+094 during 2019 April to be "faint" and, yet, the variation observed was inherent rather than due to the nearby sources.

\subsection{Variability study from the light curve}
We generated the light curve of 100 sec time binning of the total 40~ks data to study the variability (Fig.~\ref{src-bkg}a). The light curve 
shows a small variability during this period. We also generated light curves of 14 individual orbits of $0.01$ sec time binning. The power 
density spectra (PDS) do not show any quasi-periodic-oscillations (QPOs). This observation contrasted with the 2002 observations of the 
object where a broad QPO was observed, and the spectrum was harder \citep{Gogus04}. 

\begin{figure}[ht]
    \centering
    \includegraphics[width=10.0cm,keepaspectratio=true]{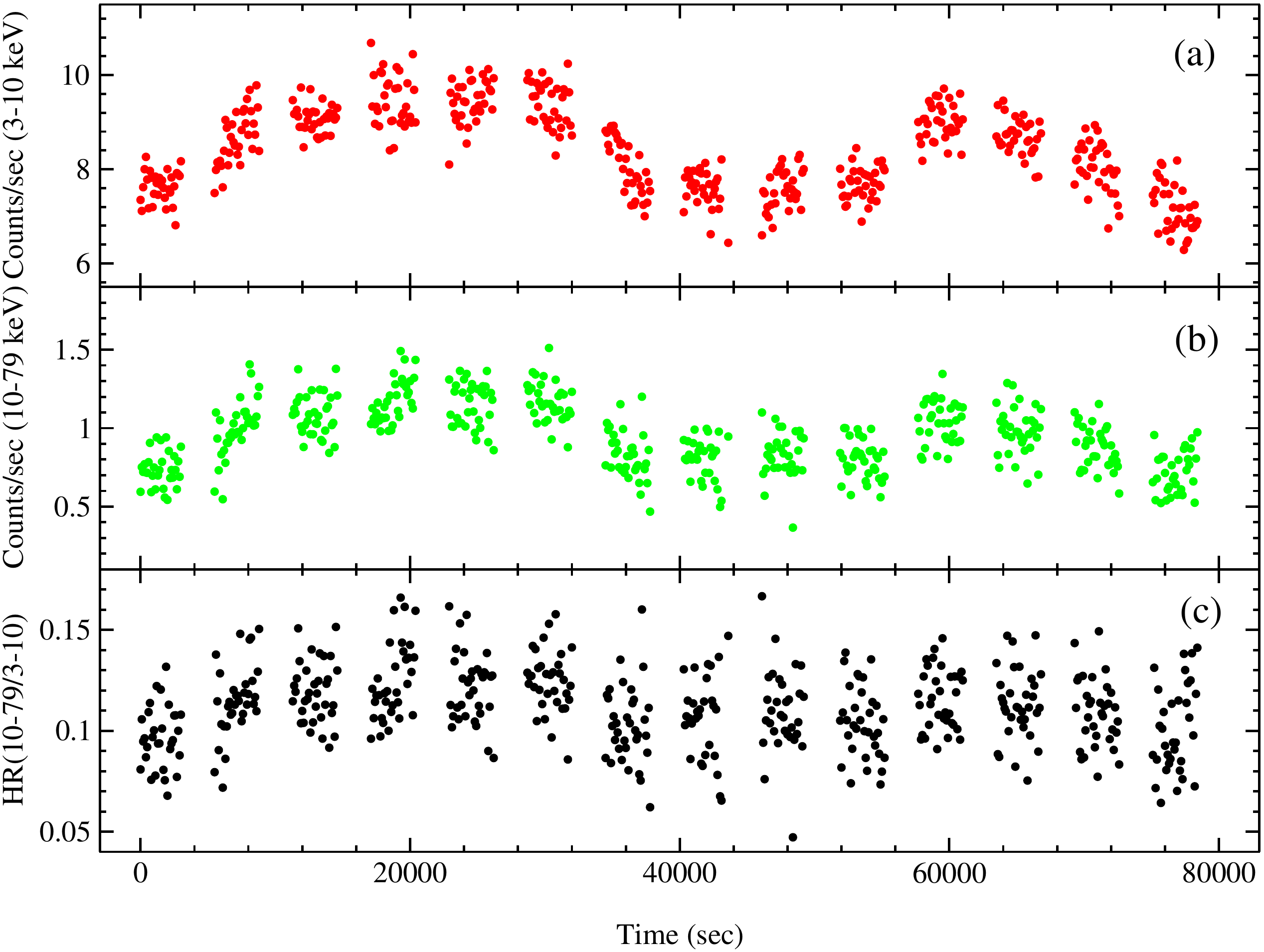}
    \caption{The variation of (a) 3-10 keV count rate, (b) 10-79 keV count rate, and (c) hardness ratio (HR) with time (in sec). The light curves 
are binned with a time resolution of 100 sec.}
    \label{lc-hr}
\end{figure}

We generated light curves of the entire observation in the energy ranges of 3-10 keV and 10-79 keV. The variation of these light curves along with the 
hardness ratio (HR; between 10-79 keV and 3-10 keV) are shown in Fig~\ref{lc-hr}. We noticed that the soft photon (3-10 keV) counts always dominate 
the hard photon (10-79 keV) counts. Both the soft and hard count rates showed coherent periodic variation during the observation. Thus the HR always 
remained at low values ($\sim$0.05 to $\sim$0.15). The soft and the hard count rates and the value of the HR indicate that the source was in the softer 
spectral regime during our studied period. 

\subsection{Accretion properties from spectral analysis}
\begin{figure}
    \centering
    \includegraphics[width=7.0cm,keepaspectratio=true]{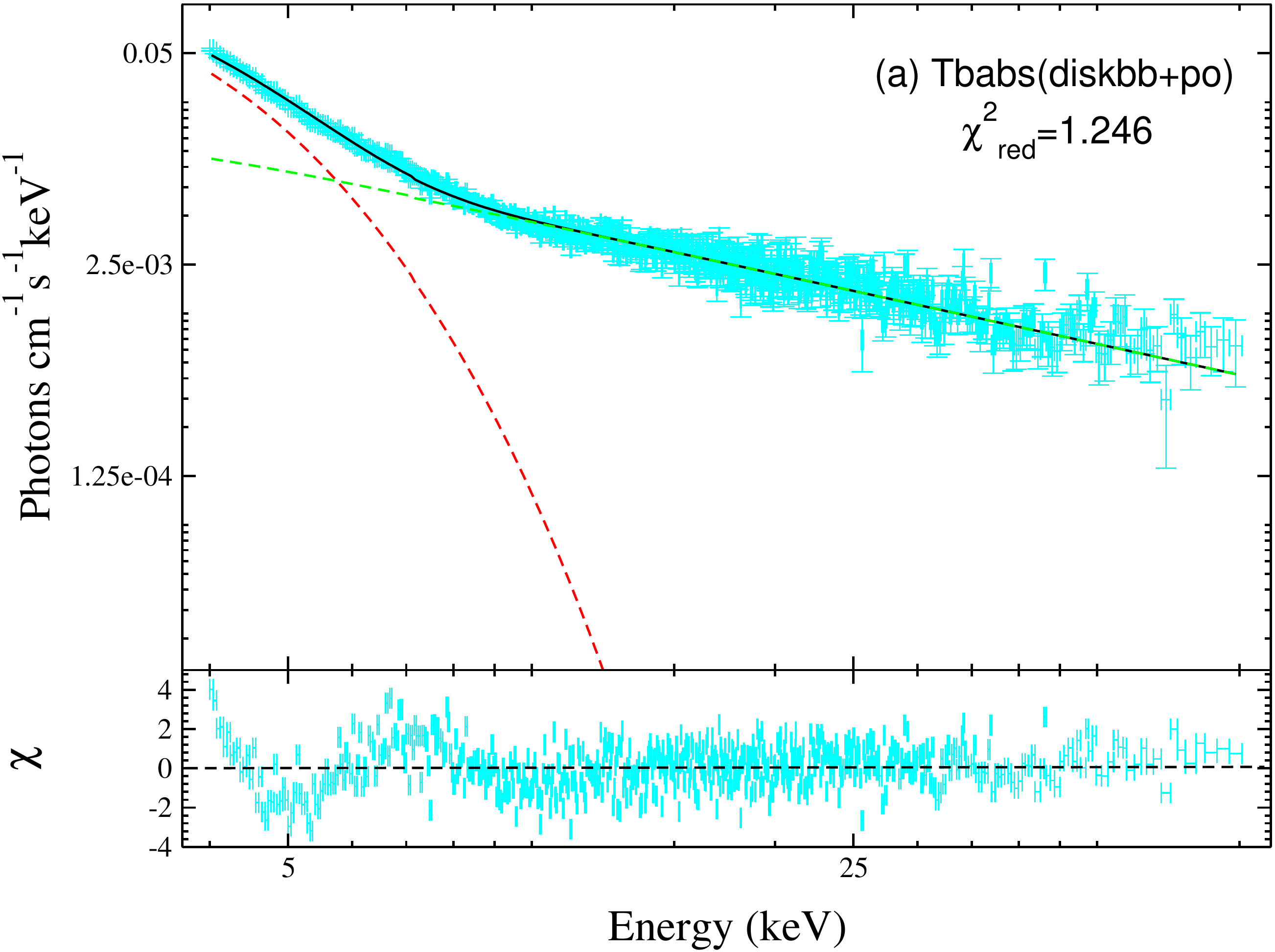}
    \includegraphics[width=7.0cm,keepaspectratio=true]{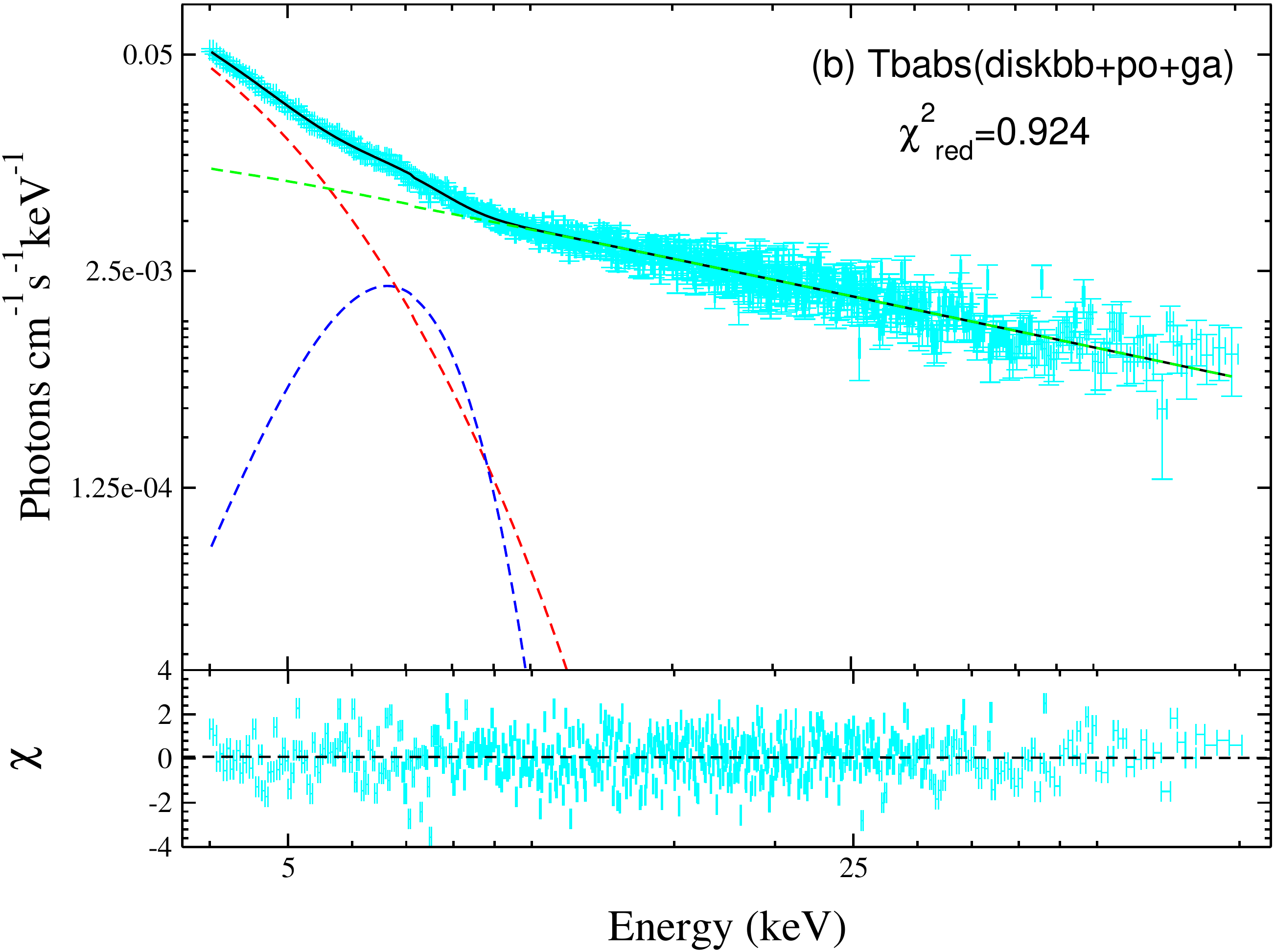}
    \includegraphics[width=7.0cm,keepaspectratio=true]{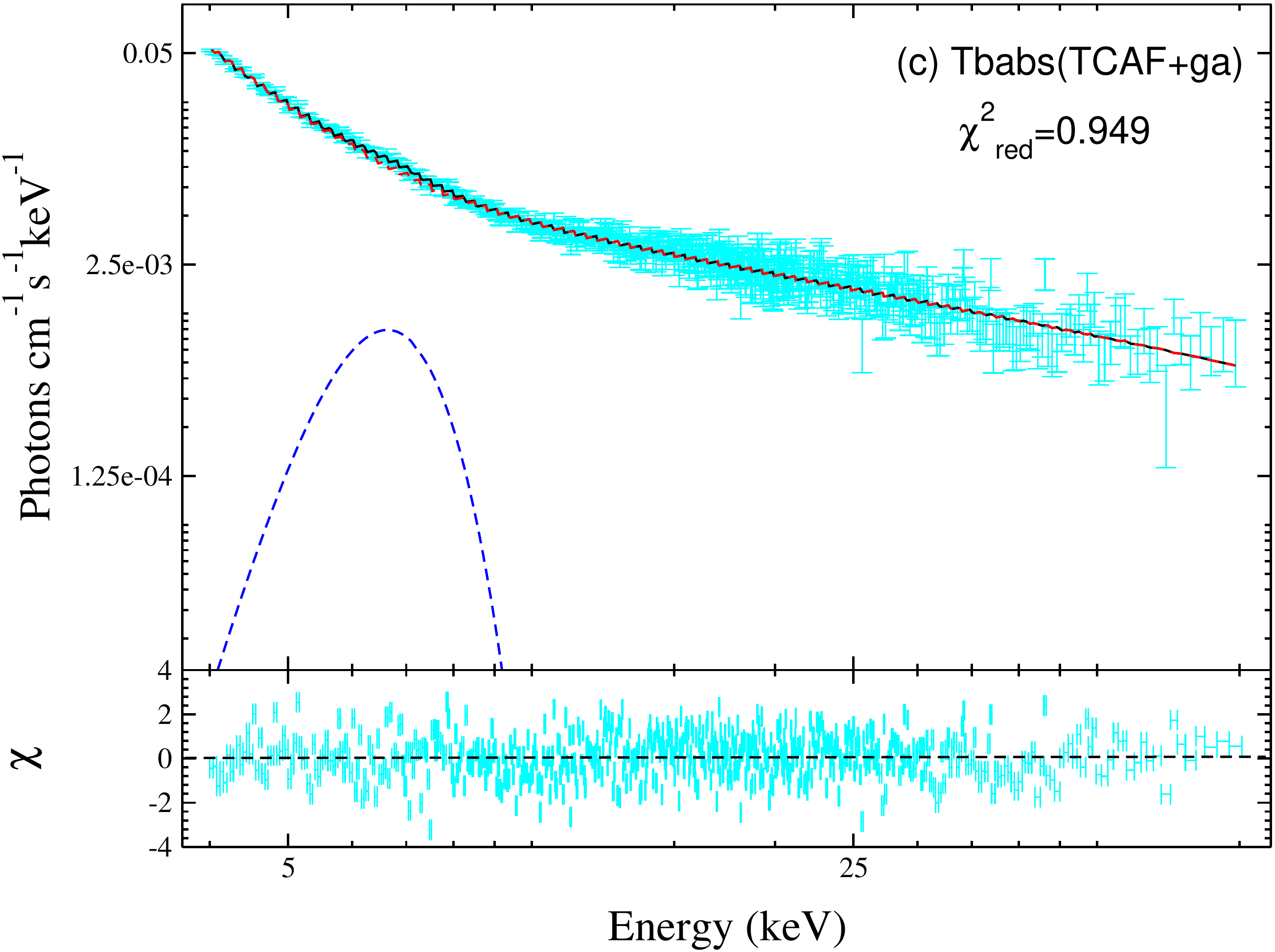}
    \caption{Unfolded spectra using (a) {\tt Tbabs(diskbb+powerlaw)}, (b) {\tt Tbabs(diskbb+powerlaw+Gaussian)}, and (c) {\tt Tbabs(TCAF+Gaussian)} with their ratio. The red, green dashed lines are the {\tt blackbody} and {\tt powerlaw} model components. The solid black line refers to the combined best fitted spectra. The blue dashed lines are the {\tt Gaussian} component.}
    \label{spec1}
\end{figure}

We studied the spectral properties with both phenomenological and physical models separately. We first fitted the spectrum of the full observation 
($\sim$40~ks) using combined {\tt diskbb} and {\tt powerlaw} models (Fig.~\ref{spec1}a). The inner disk temperature ($T_{in}$) was obtained to be 
0.779~keV. The high photon index ($\Gamma~\sim~2.085$) indicates a softer spectral state. The fitted spectrum retained a $\chi^2_{red}$ 
value of 744/597 $\sim$ 1.246. We noticed a small residual at around 6.5 keV (see, Fig.~\ref{spec1}a). To check the contribution of any Fe-K$\alpha$ 
line in the spectrum, we fitted the spectrum with {\tt Tbabs(diskbb+powerlaw+Gaussian)} model (Fig.~\ref{spec1}b). The inner disk temperature 
($T_{in}$) decreased very little (0.645~keV). The photon index also decreased to ($\Gamma~\sim~2.02$). We obtained 6.48 keV as the 
line energy and $\sim$1 keV as the line width ($\sigma$) from {\tt Gaussian} model. The best fitted spectrum gave a $\chi^2_{red}$ value of
549/594 $\sim$ 0.924. The fit improved with the inclusion of the {\tt Gaussian} model for the Fe K$\alpha$ emission line. We also fitted the 
spectrum with TCAF {\tt fits} file along with {\tt Gaussian} model (Fig.~\ref{spec1}c). The disk rate was high ($\dot m_d\sim1.699$) 
compared to the halo rate ($\dot m_h\sim0.160$). The shock location and compression value were obtained as $X_s\sim49.83~r_s$ and 
$R\sim1.1$ respectively. These accretion rates and shock parameters indicating a softer state also justify the results obtained from the 
combined {\tt diskbb+powerlaw+Gaussian} model fitting. The best-fitted spectrum retained a $\chi^2_{red}$ value of 562/592 $\sim$ 0.949. 
The best-fitted spectra using these three different combinations of models are shown in Fig.~\ref{spec1}.

\begin{specialtable}
\begin{center}
\caption{Spectral Analysis Results obtained from {\it NuSTAR} Observation}
\label{table1}
\begin{tabular}{lccc}
\hline
	Model1&Diskbb&$T_{in}$ (keV)& $0.779^{+0.013}_{-0.013}$ \\
	&&norm& $153^{+19}_{-16}$ \\
	&Powerlaw& $\Gamma$& $2.085^{+0.028}_{-0.028}$\\
	&& norm & $0.057^{+0.004}_{-0.004}$\\
	&&$\chi^2$/dof& 744/597 $\sim$ 1.246\\
\hline
	Model2&Diskbb & $T_{in}$ (keV)& $0.645^{+0.019}_{-0.019}$ \\
	&& norm & $493^{+95}_{-72}$ \\
	&Powerlaw& $\Gamma$ & $2.023^{+0.028}_{-0.028}$ \\
	&& norm & $0.047^{+0.003}_{-0.003}$ \\
	&Gaussian &$E$ (keV)& $6.483^{+0.152}_{-0.147}$ \\
	&& $\sigma$ (keV) & $1.000_{-0.041}$\\
	&& norm & $9e^{-4}\pm1e^{-4}$\\
	&& $\chi^2$/dof& 549/594 $\sim$ 0.924 \\
\hline
	Model3&TCAF & $\dot{m}_d$ ($\dot{M}_{Edd}$)&$1.699^{+0.03}_{-0.04}$\\
	&& $\dot{m}_h$ ($\dot{M}_{Edd}$)&$0.160^{+0.03}_{-0.04}$ \\
	&& $X_s$ ($r_s$)&$49.831^{+9}_{-11}$ \\
	&& $R$ & $1.100^{+0.15}_{-0.16}$ \\
	&& $M_{BH}$ ($M_\odot$)& $6.516^{+0.53}_{-0.76}$ \\
	&& $N_{TCAF}$& $312^{+0.08}_{-0.11}$ \\
	&Gaussian &$E$ (keV)& $6.521^{+0.183}_{-0.175}$ \\
	&& $\sigma$ (keV) & $0.814^{+0.045}_{-0.151}$\\
	&& norm & $5e^{-4}\pm1e^{-4}$\\
	&& $\chi^2$/dof&562/592 $\sim$ 0.949 \\
\hline
\end{tabular} 
\end{center}
\end{specialtable}

\begin{figure}
    \includegraphics[width=6.5cm,keepaspectratio=true]{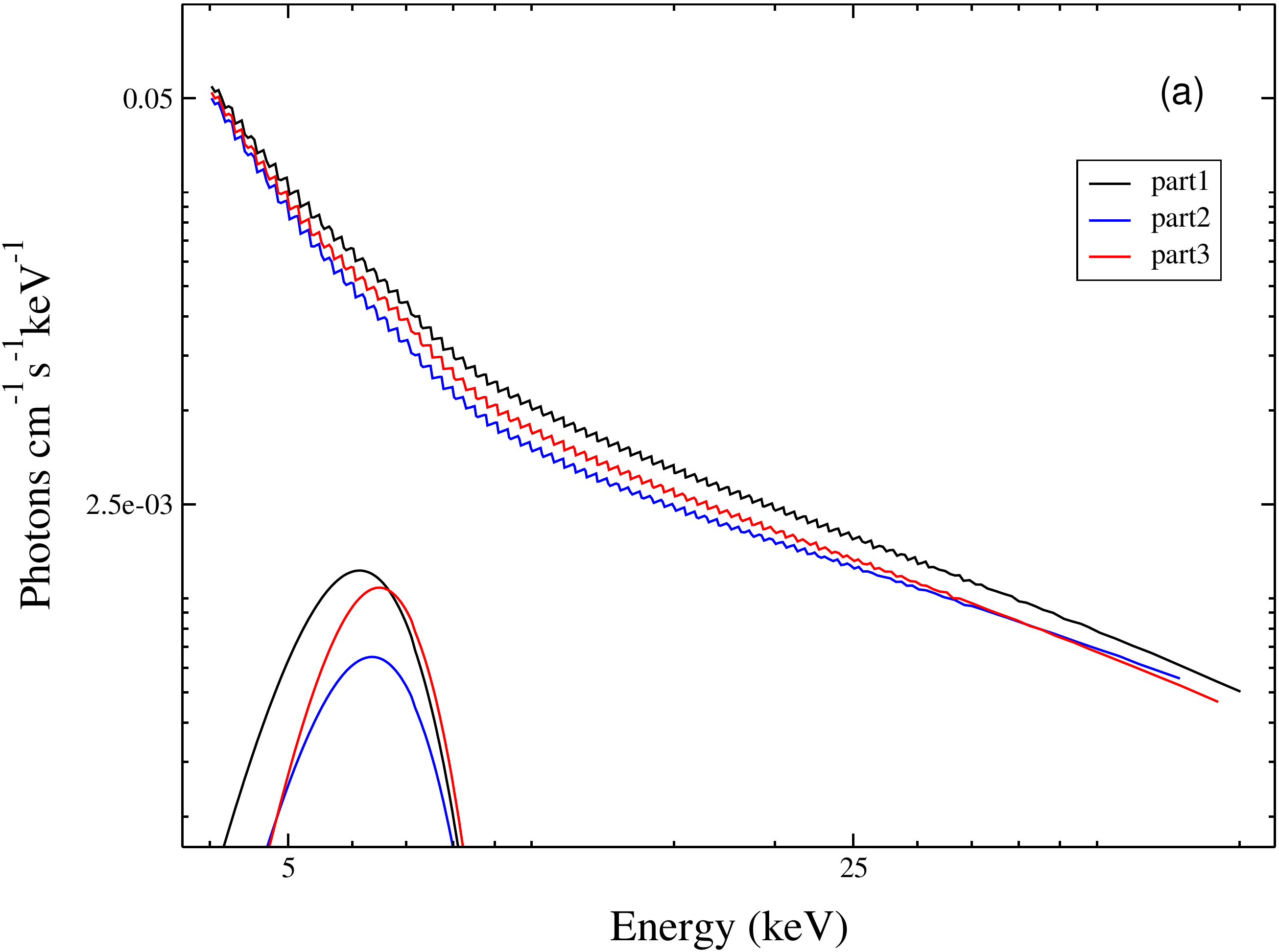}
    \hspace{0.3cm}
    \includegraphics[width=6.1cm,keepaspectratio=true]{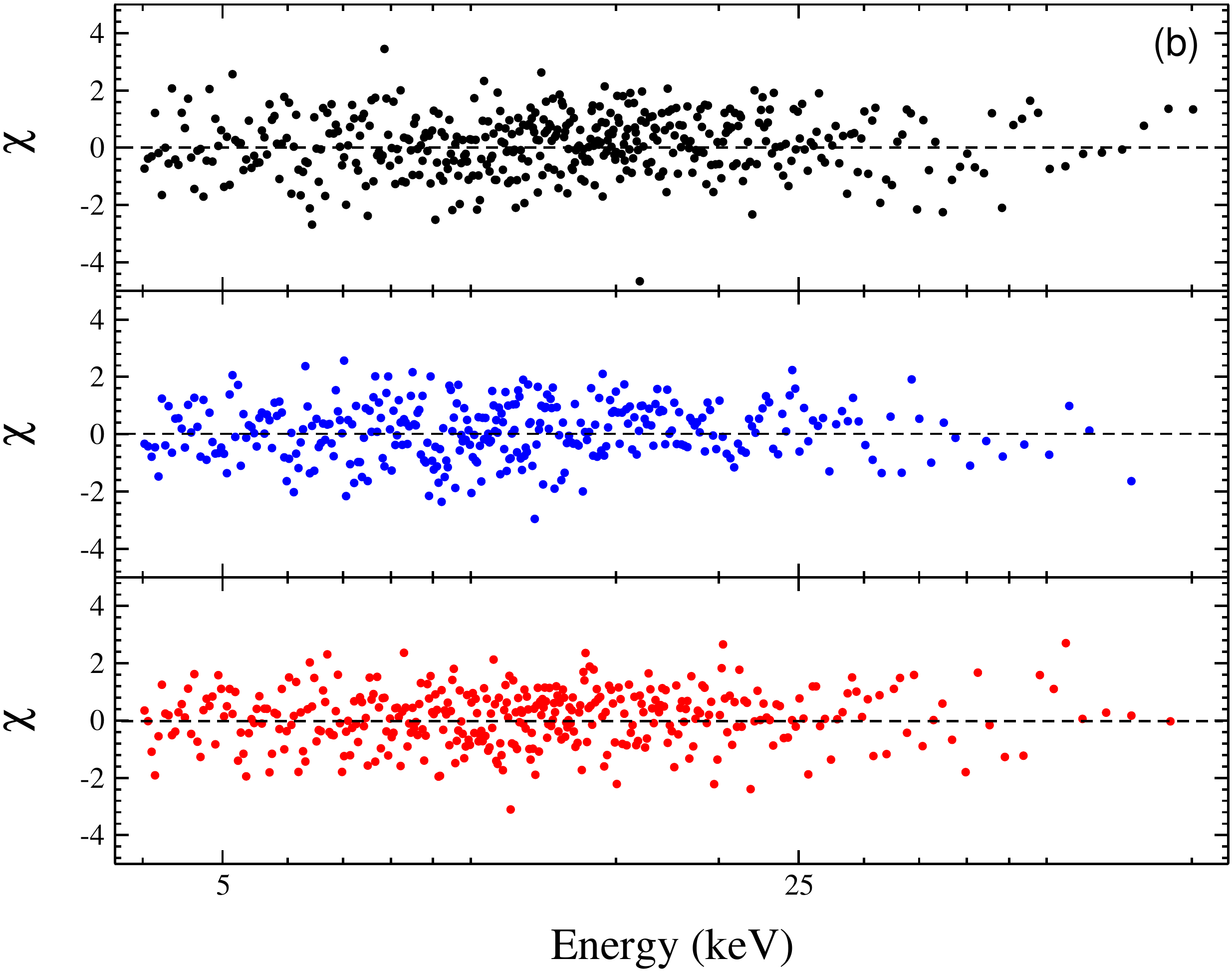}
    \caption{(a) Best-fitted spectra for the three segments of the {\it NuSTAR} observation. The right panel (b) shows the $\chi$ (i.e. $\frac{\rm data-\rm model}{\rm error}$) variations of the three spectra in respective colors.}
    \label{spec2}
\end{figure}

Since a variation can be easily noticed in the light curve of the full observation (Fig.~\ref{src-bkg}a), we divided the light curve into 
three segments of $\sim$16~ks, $\sim$10~ks and $\sim$12~ks and analyzed for detailed study. We fitted these three spectra using two 
combinations {\tt Tbabs(diskbb+powerlaw+Gaussian)} and {\tt Tbabs(TCAF+Gaussian)}. The spectral parameters are listed in Table~\ref{table2}. 
Only TCAF plus Gaussian model fitted spectra are shown in Fig.~\ref{spec2}.

\begin{specialtable}
\begin{center}
    \caption{TCAF model fitted spectral parameters for three segments of the {\it NuSTAR} observation}
\label{table2}
\begin{tabular}{lcccc}
\hline
	& Obs.& part1 & part2 & part3 \\
\hline
	& MJD & 58583.471 & 58583.709 & 58583.975 \\
\hline
Diskbb & $T_{in}$ (keV)& $0.652^{+0.033}_{-0.035}$ & $0.701^{+0.034}_{-0.032}$& $0.668^{+0.034}_{-0.035}$\\
	& norm & $593^{+55}_{-99}$ & $467^{+47}_{-96}$& $487^{+34}_{-51}$\\
	Powerlaw& $\Gamma$ & $2.084^{+0.043}_{-0.043}$ & $1.936^{+0.063}_{-0.063}$& $2.044^{+0.056}_{-0.055}$\\
	& norm & $0.063^{+0.007}_{-0.006}$ & $0.032^{+0.005}_{-0.005}$& $0.047^{+0.007}_{-0.006}$\\
	Gaussian& $E$ (keV)&  $6.179^{+0.255}_{-0.245}$ & $6.768^{+0.311}_{-0.338}$ & $6.450^{+0.243}_{-0.227}$\\
	& $\sigma$ (keV) &  $1.000_{-0.088}$  & $1.000_{-0.137}$ & $1.000_{-0.137}$\\
	& norm & $9e^{-4}\pm2e-4$ & $6e^{-4}\pm1e^{-4}$ & $9e^{-4}\pm2e^{-4}$\\
	& $\chi^2$/dof & 483/439  $\sim$ 1.100 & 316/332  $\sim$ 0.952 & 352/368  $\sim$ 0.956\\
\hline
	 TCAF &$\dot{m}_d$ ($\dot{M}_{Edd}$)&$1.644^{+0.03}_{-0.04}$&$1.572^{+0.03}_{-0.04}$&$1.884^{+0.03}_{-0.04}$\\
	 &$\dot{m}_h$ ($\dot{M}_{Edd}$)&$0.137^{+0.03}_{-0.04}$&$0.165^{+0.03}_{-0.04}$&$0.139^{+0.03}_{-0.04}$ \\
	 &$X_s$ ($r_s$)&$46.117^{+9}_{-11}$&$50.140^{+9}_{-11}$&$47.480^{+9}_{-11}$ \\
	 &$R$ & $1.100^{+0.15}_{-0.16}$& $1.101^{+0.15}_{-0.16}$& $1.100^{+0.15}_{-0.16}$ \\
	 &$M_{BH}$ ($M_\odot$)& $6.380^{+0.73}_{-0.66}$& $6.702^{+0.73}_{-0.66}$& $6.721^{+0.53}_{-0.76}$ \\
	 &$N_{TCAF}$& $326^{+0.08}_{-0.11}$& $227^{+0.08}_{-0.11}$& $330^{+0.08}_{-0.11}$ \\
	 Gaussian &$E$ (keV)& $5.959^{+0.297}_{-0.278}$& $6.600^{+0.310}_{-215}$& $6.348^{+0.195}_{-0.201}$ \\
	& $\sigma$ (keV) & $1.000_{-0.145}$ & $0.901^{+0.139}_{-0.220}$ & $0.900^{+0.099}_{-0.131}$\\
	& norm & $7e^{-4}\pm2e^{-4}$ & $3e^{-4}\pm03e^{-5}$ & $7e^{-4}\pm03e^{-4}$\\
	 &$\chi^2$/dof & 483/437 $\sim$ 1.105 & 321/330 $\sim$ 0.973 & 350/366  $\sim$ 0.956\\
\hline
\end{tabular} 
\end{center}
\end{specialtable}

\subsection{Estimation of Inner disk radius and Prediction of Mass}
We estimated the mass for XTE~J1908-094 from the spectral analysis. From the obtained $\chi^{2}_{red}$ values of model 1 and model 2 
in Table~\ref{table1}, we suggested model 2 is the best fitted phenomenological model. We considered the fitted disk-normalization ($N_{disk}$) 
to estimate the mass of the BHC XTE J1908+094. We average out the $N_{disk}$ values obtained from fittings of the three segments of the 
whole observation (Table~\ref{table2}). The inner radius ($r_{in}$) of the disk and the {\tt diskbb} normalization ($N_{disk}$) are related as,
\begin{equation}
    N_{disk}=\left(\frac{r_{in}}{D/10~kpc}\right)^2 \cos\theta.
\end{equation}
$D$ is the distance of the system in $kpc$ and $r_{in}$ is in $km$. $\theta$ is the inclination of the disk in degree. This estimated inner 
radius from the above equation is subjected to some errors (see, \cite{Shimura95,Kubota98}). The corrected inner radius ($R_{in}$ ($km$)) is,
\begin{equation}
    R_{in} (km) \simeq \kappa^2\xi r_{in}.
\end{equation}
$\kappa$ and $\xi$ are the hardening factor \citep{Shimura95} and inner boundary correction factor \citep{Kubota98} respectively. 
Since, the inclination angle of the system is not confirmed, we considered three guess values for the inclination as: $\theta$ $\sim$ $30^\circ$, 
$50^\circ$, and $80^\circ$. The average value of $N_{disk}$ was 516. We consider the  $\kappa$ and $\xi$ are 1.7 and 0.41 respectively (\cite{Shimura95,Kubota98}) and assume the distance of the system as 10 $kpc$. We obtained the inner radius ($R_{in}$) for to be $29~km$, $34~km$, 
and $65~km$, for the inclination angle $30^\circ$, $50^\circ$, and $80^\circ$, respectively. The inner edge of the disk ($R_{in}$) is considered 
to be truncated at the inner most stable circular orbit (ISCO). For a Schwarzschild black hole it is $6GM_{BH}/c^2$, where $G$, $M_{BH}$ and 
$c$ are the Gravitational constant, mass of the black hole and speed of light at vacuum. We obtained the mass for this black hole from the 
obtained $R_{in}$ values to be 3.2, 3.7, and 7.2 $M_\odot$ respectively.

In TCAF, the mass of the BH is an important input parameter. If it is not well known from dynamical or other methods, one can get the best-fitted 
value of the $M_{BH}$ from each spectral fit. Here, we obtained the mass of XTE~J1908+094 as $\sim$6.5 $M_\odot$ from the combined TCAF, and 
Gaussian model fit when the entire duration of the observation was considered. (Table~\ref{table1}). 

\section{Conclusions}
We studied the spectral and temporal properties of BHC XTE~J1908+094 during its renewed X-ray activity using one {\it NuSTAR} observation in 2019, 
April 10. The X-ray activity was very faint and mostly dominated by soft photons. No quasi-periodic oscillations (QPOs) were observed in its power 
density spectra (PDS). The light curves generated in the soft (3-10 keV) and hard (10-79 keV) energy range showed coherent periodic variation. 
The low value of hardness ratio suggests that the source was in a softer spectral state during the observation. 

We analyzed the spectrum using both phenomenological and physical models separately. The high power-law photon index ($\Gamma\sim 2.23$) 
indicates that the source was either in soft intermediate or in soft state. The disk normalization was also relatively high compared to the 
power-law normalization. Due to the small residual around 6.5 keV, we refitted the spectrum, adding a {\tt Gaussian} for the contribution of 
Fe K$\alpha$ line emission. A broad {\tt Gaussian} was obtained with line energy $\sim 6.48~keV$ and line width ($\sigma$) $\sim1~keV$. 
The photon index ($\Gamma$) decreased to $2.02$, suggesting that the source probably was in the soft intermediate state.  The presence of 
broad width ($\sigma \ge 1$) Gaussian line also signifies the spectral state as soft intermediate. While fitting the three segments of the 
whole observation divided due to the periodic variability, we noticed that the second segment (see, Table~\ref{table2}) showed a little 
harder spectrum than the other two segments (part1 and part3).

We refitted the spectrum using TCAF plus Gaussian model. We noticed an unusually high disk rate ($\dot{m}_d\sim1.699~\dot{M}_{Edd}$) over 
the halo rate ($\dot{m}_h\sim0.160~\dot{M}_{Edd}$). This high amount of hard photon contribution is also noticed in the variation of 
light curves (see, Fig.~\ref{lc-hr}). The obtained shock location ($X_s\sim49.83~r_s$) and compression ratio ($R\sim1.1$) also refer to 
soft intermediate spectral state. The spectral parameters obtained from the whole observation's partial fitting also suggest that 
during the mid-segment, when both soft and hard count decreased, the spectrum became harder but remained in the same spectral state. 
This faint X-ray activity of XTE~J1908+094 in 2019 could be due to the sudden enhancement of viscosity and supply of matter from the pile-up radius. 
This type of behavior has been observed in the case of H1743-322 \citep{SKC19} and GX~339-4 \citep{RB20}. As the supply of residual matter from 
the previous outburst (2012-2013) was exhausted, the very short-term X-ray activity was also faded.

We also estimated the possible value of the mass of the black hole ($M_{BH}$) from the spectral analysis. During our studied period, the spectra 
are dominated by the soft photons coming from the disk. The inner radius can be constrained in the soft spectral state from the {\tt diskbb} 
normalization ($N_{disk}$). Since the other two unknown variables (distance and inclination) are not confirmed for this source, we consider the 
distance to be 10~kpc and inclination as 30$^\circ$, 50$^\circ$ and 80$^\circ$. The corrected inner radius ($R_{in}$) was obtained as 29 km, 34 km, 
and 65 km. As the source is found in the soft-intermediate state, the inner radius must be located very close to the ISCO \citep{Done07}. Considering 
a Schwarzschild black hole the ISCO is at $\sim \frac{6GM_{BH}}{c^2}$ or $3~r_s$. We obtained the black hole mass to be 3.2, 3.7 and 7.2 $M_\odot$ 
for those three $R_{in}$. This result would only be valid for a Schwarzschild black hole. For a Kerr black hole, the inner radius would be located 
at less than $3~r_s$, affecting the prediction of mass from the disk normalization. We could suggest that the system is located at $\sim 8-10~kpc$ 
and is highly inclined ($>70^\circ$). From the TCAF model fitted spectral fit, we also obtained the mass of the black hole to be $\sim$6.5 $M_\odot$.

\section{Summary}
We studied the faint outburst of XTE~J1908+094 in 2019 using archival {\it NuSTAR} data. From the timing and spectral study, we conclude that:\\
i) No quasi-periodic oscillation (QPO) was found in the PDS.\\
ii) The source was in SIMS during our studied period. \\
iii) We also estimated the most probable mass of the black hole to be $\sim$6.5 $M_\odot$.


\vspace{6pt}


\dataavailability{The data used here are publicly available. This research has made use of the NuSTAR Data Analysis Software (NuSTARDAS) jointly developed by the ASI Science Data Center (ASDC, Italy) and the California Institute of Technology (USA).}

\acknowledgments{D.C. acknowledges the support of the PDR fellowship, IIA, Bengaluru, Karnataka, India..
A.J. acknowledges the support of the Post-Doctoral Fellowship from Physical Research Laboratory, Ahmedabad, India, funded by the
Department of Space, Government of India. 
K.C. acknowledges support from DST/INSPIRE (IF170233) fellowship.
R.B. acknowledges support from CSIR-UGC NET qualified UGC fellowship (June-2018, 527223).
S.N., D.D. and S.K.C. acknowledge support ISRO sponsored RESPOND project (ISRO/RES/2/418/17-18) fund.
D.D. and S.K.C. also acknowledge support from Govt. of West Bengal, India and DST/GITA sponsored India-Taiwan collaborative project (GITA/DST/TWN/P-76/2017) fund.}


\end{paracol}
\reftitle{References}

\end{document}